\documentclass[preprint,pra,floatfix]{revtex4}

\usepackage{graphicx}

\begin{document}

\title{WavePacket: A Matlab package for numerical quantum dynamics.\\
III: Quantum-classical simulations and surface hopping trajectories} 

\author{Burkhard Schmidt}
\email{burkhard.schmidt@fu-berlin.de}
\affiliation{
Institut f\"{u}r Mathematik, Freie Universit\"{a}t Berlin\\ Arnimallee~6, D-14195 Berlin, Germany}
\author{Rupert Klein}
\email{rupert.klein@fu-berlin.de}
\affiliation{
Institut f\"{u}r Mathematik, Freie Universit\"{a}t Berlin\\ Arnimallee~6, D-14195 Berlin, Germany}
\author{Leonardo Cancissu Araujo}
\email{leonardo.araujo@tum.de}
\affiliation{Zentrum Mathematik, Technische Universit\"{a}t München\\ Boltzmannstr.~3, D-85747 Garching, Germany}

\date{\today}

\begin{abstract}
WavePacket is an open-source program package for numerical simulations in quantum dynamics. 
It can solve time-independent or time-dependent linear Schr\"{o}dinger and Liouville-von Neumann-equations in one or more dimensions. 
Also coupled equations can be treated, which allows, e.g., to simulate molecular quantum dynamics beyond the Born-Oppenheimer approximation.
Optionally accounting for the interaction with external electric fields within the semi-classical dipole approximation, WavePacket can be used to simulate experiments involving tailored light pulses in photo-induced physics or chemistry.
Being highly versatile and offering visualization of quantum dynamics 'on the fly', WavePacket is well suited for teaching or research projects in atomic, molecular and optical physics as well as in physical or theoretical chemistry.

Building on the previous Part~I [Comp. Phys. Comm. \textbf{213}, 223-234 (2017)] and Part~II [Comp. Phys. Comm. \textbf{228}, 229-244 (2018)] which dealt with quantum dynamics of closed and open systems, respectively, the present Part~III adds fully classical and mixed quantum-classical propagations to WavePacket.
In those simulations classical phase-space densities are sampled by trajectories which follow (diabatic or adiabatic) potential energy surfaces.
In the vicinity of (genuine or avoided) intersections of those surfaces trajectories may switch between surfaces.
To model these transitions, two classes of stochastic algorithms have been implemented: (1) J. C. Tully's fewest switches surface hopping and (2) Landau-Zener based single switch surface hopping.
The latter one offers the advantage of being based on adiabatic energy gaps only, thus not requiring non-adiabatic coupling information any more. 

The present work describes the MATLAB version of WavePacket 6.0.2 which is essentially an object-oriented rewrite of previous versions, allowing to perform fully classical, quantum--classical and quantum-mechanical simulations on an equal footing, i.~e., for the same physical system described by the same WavePacket input.
The software package is hosted and further developed at the Sourceforge platform, where also extensive Wiki-documentation as well as numerous worked-out demonstration examples with animated graphics are available. 
\end{abstract}

\maketitle


\section{Introduction}
\label{sec:intro}

Progress in the generation of short intense laser pulses and related experimental techniques on an ultra-fast time scale has lead to substantial advances in atomic and molecular physics and related fields in the late 20th century~\cite{Zewail2000}.
This has also motivated new developments in theoretical and simulation studies of quantum molecular dynamics in recent years~\cite{May2000,Tannor2004}. 
However, despite the obvious need, general-purpose and freely available simulation software in this field is still scarce.
Among the exceptions, there is the versatile MCTDH package which has evolved into a quasi-standard in quantum molecular dynamics~\cite{Beck2000}.
In addition, there is also the TDDVR package in that field~\cite{Khan2014}.
Software packages more commonly used in the physics community include QuTiP for the dynamics of open quantum systems~\cite{Johansson2013}, the FermiFab toolbox for many-particle quantum systems~\cite{Mendl2011}, and the QLib platform for numerical optimal control~\cite{Machnes2011}.
The present article deals with the \textsc{WavePacket} software package.
Its main version which is coded in \textsc{Matlab} has been described in a series of two recent articles~\cite{WavePacket1,WavePacket2}. 
Part I focuses on closed quantum systems and the solution of Schr\"{o}dinger equations, with emphasis on discrete variable representations (DVR), finite basis representations (FBR), and various techniques for temporal discretization~\cite{WavePacket1}.
Part II is mainly on open quantum systems and the solution of Liouville--von Neumann equations, optimal control of quantum systems and their dimension reduction~\cite {WavePacket2}.
It is emphasized that the target systems for \textsc{WavePacket} are low- to medium-dimensional (model) systems where computational requirements are not the dominant concern.
Instead, the user-friendliness of the \textsc{Matlab} environment and, in particular, the ability of generating on-the-fly graphics have attracted an increasing number of users.
As such, \textsc{WavePacket} is very suitable not only for educational purposes, but also for development, implementation and testing of various numerical techniques and algorithms which is facilitated by the highly modular structure of the software package. 

While fully quantum-dynamical simulations are nowadays routinely carried out for small molecules, the treatment of larger systems such as atomic and molecular clusters, biologically relevant molecules, and condensed matter systems remains a challenge due to the high computational effort. 
Despite impressive progress in numerical quantum dynamics, especially by the multi-layer extensions of the MCTDH~\cite{Wang2003} methodology, there is still the need for more approximate classical or mixed (hybrid) quantum-classical computational methods.
Those represent an important alternative not only because their computational effort scales more favorably with the system size, but often also because they provide intuitive insight into the dynamics of molecular processes.
Among these approaches is the surface hopping trajectory (SHT) simulation technique.
The basic idea of SHT is to propagate classical trajectories for the heavy particles (typically nuclei) which may statistically hop between different quantum states of the light particles (typically electrons) subsystem thus modeling nonadiabatic transitions in a simple way.
Even though the seminal paper by J.~C.~Tully on the \textsl{fewest switches surface hopping} (FSSH) algorithm was published almost 30 years ago~\cite{Tully1990} this method is still in active use, due to its low computational expense and its extremely simple implementation.
In the chemical community, this technique has become routinely available through its implementation in software packages such as Newton-X~\cite{Barbatti2014}, Fish~\cite{Mitric2009}, Sharc~\cite{Richter2011,Mai2018} providing a combination of SHT techniques with standard electronic structure software packages.
In those approaches, the forces and the nonadiabatic couplings governing the trajectories are computed “on-the-fly” by means of \textsl{ab initio} or semi-empirical electronic-structure calculations.

Moreover, there has been substantial methodological progress in SHT simulations in recent years~\cite{Tully2012,Wang2016}.
On the one hand, these efforts aim at overcoming the known problems of the earlier SHT approaches. 
Among others, the lack of communication between the evolving trajectories leads to overcoherence, and limitations in the energy conservation are hampering a description of superexchange processes~\cite{Subotnik2016,Wang2016}. 
On the other hand, also the SHT algorithm itself has been improved. 
In the so-called \textsl{single switch surface hopping} (SSSH) approach there is only a single switch decision required each time a trajectory passes a critical region, typically a (genuine or avoided) crossing seam or conical intersection.
The transition probabilities in these SSSH algorithms are calculated from Landau-Zener (LZ) formulae~\cite {Lasser2007,Lasser2008,Fermanian2008}.
Of particular interest is a variant which is entirely based on adiabatic energy gaps, thus rendering the need for nonadiabatic coupling information completely redundant~\cite{Belyaev2011,Belyaev2014,Belyaev2015}.

In this work, we present the implementation of classical trajectory and SHT propagation techniques (both FSSH and SSSH variants) into the most recent \textsc{Matlab} version of \textsc{WavePacket} 6.0.2.
This has been made possible through an object-oriented rewrite of previous versions of \textsc{WavePacket} which were described in  Part~I~\cite{WavePacket1} and Part~II~\cite{WavePacket2}.
The main goal of using such a programming technique is to perform fully classical, mixed quantum-classical and fully quantum-mechanical simulations on an equal footing.
This allows for a direct comparison of the respective evolutions for the same physical system, i.~e., for the same Hamiltonian,  initial conditions, time stepping etc. which can be defined in the same \textsc{WavePacket} input file.  
Such a comparison strongly benefits from the generation of graphical output which has always been one of the key advantages of the \textsc{Matlab} version of \textsc{WavePacket}.
Quantum and (quantum-)classical propagations are visualized in the same way, with a large variety of different options such as curve plots, contour plots, surface plots, etc. available, which helps the user to develop a more intuitive understanding of the respective type of dynamics.

In addition to the mature \textsc{Matlab} version of \textsc{WavePacket} presented here, there is also a C++ version which is however still in a very early stage of development. Both versions are hosted and further developed at the open source \textsc{SourceForge} platform where also extensive Wiki documentation as well as a large number of demonstration examples can be found.

\section{WavePacket workflow}
\label{sec:mat}
A typical workflow for a dynamical \textsc{WavePacket} simulation could be as follows
{\small\begin{verbatim}
qm_setup(); 
state=wave(); | state=traj();
qm_init(state); 
qm_propa(state); 
qm_cleanup();
\end{verbatim}}
\noindent After calling the function \texttt{qm\_setup} which opens the logfile and purges the workspace from previous calculations, the second command creates an object named \texttt{state}.
This object can be an instance of either one of the following two classes:  
Class \texttt{wave} is meant for fully quantum-mechanical simulations dealing with wavefunctions represented on grids. 
In contrast, class \texttt{traj} is designed for fully classical or hybrid quantum-classical simulations, based on classical densities sampled by swarms of trajectories. 
Note that such objects were not yet in use in version 5 described in Part~I and Part~II.
Once being constructed, these objects may be modified within the initialization function \texttt{qm\_init} which is intended to set many parameters defining the physical system and which has to be provided by the user separately for every simulation; for more information, see Sec.~\ref{sec:init}.
Next, the object \texttt{state} is propagated in time by calling the function \texttt{qm\_propa} which represents the main workhorse of the \textsc{WavePacket} software package.
For in-depth explanation of quantum-mechanical or quantum-classical propagations, see Sec.~\ref{sec:q_m} or Sec.~\ref{sec:q_c}, respectively.
Finally, the function \texttt{qm\_cleanup} closes the logfile and does other minor  cleanup.
Note that in the \textsc{Matlab} script given above the function \texttt{qm\_propa} could be replaced by \texttt{qm\_bound} which performs a bound state calculation instead of a propagation, see Sec. 5 of Part I, which is available for objects of type \texttt{wave} only.
Another alternative would be to replace the function \texttt{qm\_propa} by \texttt{qm\_movie} in which case no propagation is carried out but animated graphics is created from previously generated simulation data, see Sec. 6 of Part I.

In general, constructor methods in \textsc{Matlab} can accept input arguments, typically used to assign the data stored in properties and return initialized objects.
While in the second line of the sample script above, the constructors are called without passing any arguments, additional arguments may be passed when creating an instance of class \texttt{traj}.
In the following example
{\small\begin{verbatim}
state = traj (10000, 42);
\end{verbatim}}
\noindent an object encompassing 10000 trajectories is created. 
When the first parameter is not specified, a default value (1000 trajectories) will be assumed. 
The second parameter is used to seed the process of generation of pseudo-random numbers to ensure a predictable sequence of random numbers which may be useful for testing purposes. 
Note that if the seed is not set, a different sequence will be used in every propagation. 

After a propagation with \texttt{qm\_propa} has been carried out, calculated data is still available until the next purge by \texttt{qm\_setup}. 
For example, the global variables \texttt{time} and \texttt{expect} holding the time stepping information and all expectation values, respectively, can be imported into the current workspace with the \textsc{Matlab} declaration \texttt{global time expect}.
This information can be used, e.~g., to display the populations as a function of the time given for the discretization points of the main temporal grid, see Sec.~\ref{sec:init}. 

In summary, the rationale behind the object-oriented rewrite leading to version 6 of the \textsc{WavePacket} software package is that it is now easily possible to compare fully quantum versus quantum-classical versus fully classical dynamics for exactly the same physical system (kinetic and potential energy, initial conditions, time stepping, etc.), specified by the same initialization file \texttt{qm\_init.m}. 
Obviously, this goal is reached by polymorphism in the class definitions of \texttt{wave} and \texttt{traj}: In addition to containing all necessary data for the wave functions or trajectory bundles, repectively, these classes have to contain methods for setting up the initial conditions, the system's Hamiltonian, and its application to the system's state.
Other implemented methods deal with the propagation itself as well as with the extraction of expectation values of observables.  

Finally, it is mentioned that all the class definitions used throughout \textsc{WavePacket} 6 are realized as handle classes. 
In \textsc{Matlab} handle class constructors return handle objects, i.~e., references to the object created. 
When passing such objects to functions, \textsc{Matlab} does not have to make copies of the original objects, and functions that modify handle objects passed as input arguments do not have to return them.

\section{Initialization of WavePacket}
\label{sec:init}

A closed, non-relativistic quantum mechanical system is characterized by a Hamiltonian operator 
\begin{equation}
\hat{H}(R,-i\nabla_R,t) = \hat{T}(R,-i\nabla_R,t) + \hat{V}(R)  
\label{eq:ham}
\end{equation}
where $R$ is a position vector, $-i\nabla_R$ the corresponding momentum operator, and $T$ and $V$ are the kinetic and potential energy.
Throughout the \textsc{WavePacket} software package, atomic units are used, i.~e., Planck's constant $\hbar$, the electronic mass and the elementary charge are scaled to unity. 
Semi-classical extensions of the Hamiltonian to include the coupling to external fields shall not be treated here; for more information on this the reader is referred to Parts~I and II.
The same holds for the use of negative imaginary potentials used to absorb densities near the edges of the domain.

\begin{figure}
\begin{centering}
\includegraphics[width=0.8\textwidth]{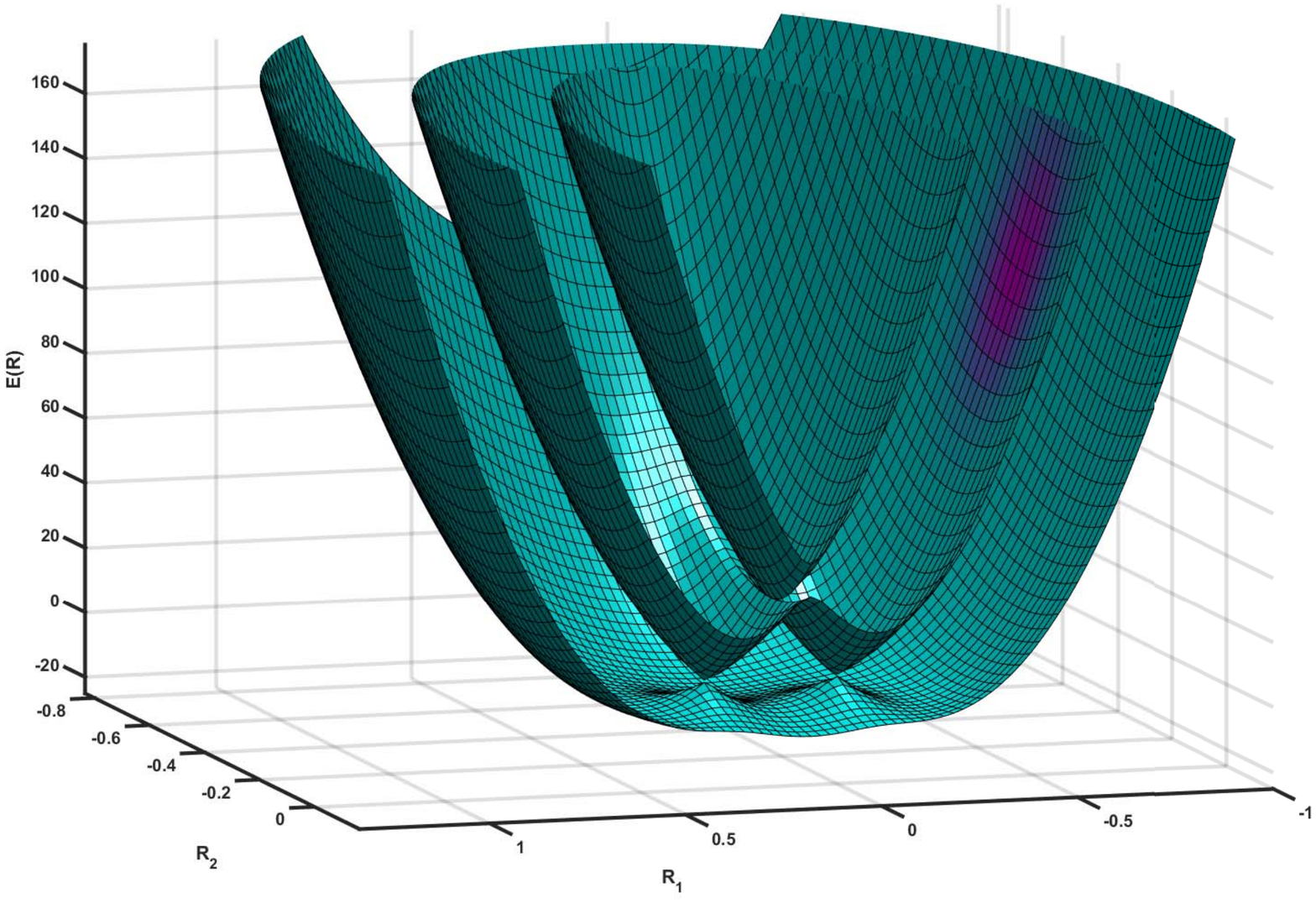}
\end{centering}
\caption{Adiabatic potential energy surfaces obtained from the Hamiltonian of Eq. (\ref{eq:JahnTeller}) displaying three conical intersections.
The blue-purple shading on the second surface indicates the initial density.
Generated with \textsc{WavePacket} option: \texttt{plots.density = vis.surface}}
\label{fig:Surfaces}
\end{figure}

Within the context of the present work it is of crucial importance that \textsc{WavePacket} can be employed not only for a single ($\nu=1$) but also for several ($\nu>1$) coupled Schr\"{o}dinger equations in which case the Hamiltonian becomes a $\nu\times\nu$ operator matrix. 
The latter case arises naturally within the field of molecular quantum dynamics, where typically $R$ specifies the nuclear degrees of freedom and where $\nu$ is the number of electronic states involved in a close coupling calculation.
Throughout this work, we will consider a prototypical example system with two spatial dimensions and with $\nu=3$ coupled channels.
The diabatic representation of its Hamiltonian is given by
\begin{eqnarray}
{\mathbf H}^\mathrm{(dia)}&=&-\frac{1}{2M}\left( \frac{\partial^2}{\partial R_1^2} + \frac{\partial^2}{\partial R_2^2} \right) \mathbf{1}_{3\times 3} \nonumber\\ 
&+& \frac{1}{2}K\left( 
	\begin{array}{ccc}
		(R_1+1/6)^2 + R_2^2 & 0 & 0 \\ 0 & (R_1-1/6)^2+R_2^2 & 0 \\ 0 & 0 & (R_1-1/2)^2+R_2^2
	\end{array}
	\right) \nonumber \\
	&+& \kappa\left(
	\begin{array}{ccc}
		0 & R_2 & 0 \\ R_2 & 0 & R_2 \\ 0 & R_2 & 0
	\end{array}\right)
	\label{eq:JahnTeller}
\end{eqnarray}
with mass $M=100$, force constant $K=600$, and coupling constant $\kappa=100$. 
This Hamiltonian represents a generalization of the two--state Jahn-Teller Hamiltonian of Ref.~\cite{Belyaev2014} to $\nu=3$, here with a value of $\epsilon=0.01$ for the quantum-classical smallness parameter which is typical for molecular systems.
The eigenvalues of the (real symmetric) potential energy matrix yield the corresponding adiabatic potential energy surfaces displaying conical intersections at $R=(0,0)$, $R=(1/6,0)$, and $R=(1/3, 0)$, see also Fig.~\ref{fig:Surfaces}.

All specifications of the above Hamiltonian, as well as further \textsc{WavePacket} settings explained below, have to be made by the user. 
This can be achieved with a user-defined function, which we suggest to call \texttt{qm\_init}. 
Typically, this function begins as follows
{\small\begin{verbatim}
function qm_init (state)
global hamilt plots space time
\end{verbatim}}
\noindent The second line serves to declare the most important variables inside \textsc{WavePacket} globally accessible. 
Note that it is a general policy throughout the \textsc{Matlab} version of \textsc{WavePacket} to use few, but highly structured variables to simplify book-keeping of
variable names. 

Subsequently, the spatial discretization has to be specified. 
Such grids are an essential ingredient of quantum-mechanical propagations of wavefunctions using objects of class \texttt{wave}, see Sec.~III C of Part~I. 
However, they also have to be specified for purely classical or quantum-classical propagations of trajectories using objects of class \texttt{traj} where they are used for graphical histogram representations of trajectory data. 
This guarantees similar appearance of graphical output, thus facilitating direct comparisons of quantum versus classical or quantum-classical dynamics.
For the example of Eq.~(\ref{eq:JahnTeller}), the ``tuning coordinate'' $R_1$ is specified by an object of class \texttt{fft} (stored in folder \texttt{+grid})
{\small\begin{verbatim}
space.dof{1}       = grid.fft;    
space.dof{1}.mass  = 100;         
space.dof{1}.n_pts = 192;         
space.dof{1}.x_min = -7/6;         
space.dof{1}.x_max = +3/2;        
\end{verbatim}}
\noindent Similarly, the spatial discretization of the ``coupling coordinate'' $R_2$ is also given by an object of class \texttt{fft}
{\small\begin{verbatim}
space.dof{2}       = grid.fft;           
space.dof{2}.mass  = 100;         
space.dof{2}.n_pts = 192;         
space.dof{2}.x_min = -2/3;         
space.dof{2}.x_max = +2/3;        
\end{verbatim}}
\noindent Here both coordinates are discretized using equally spaced grids allowing the use of FFT--methods when evaluating the kinetic operator. 
The number of points as well as the lower and upper boundaries are specified by the class properties \texttt{n\_pts}, \texttt{x\_min}, and \texttt{x\_max}, respectively.
Other discrete variable representations (DVRs), along with corresponding finite basis representations (FBRs) currently available in \textsc{WavePacket} are the Gauss--Legendre and Gauss-Hermite schemes, see Part~I, which, however, are not yet available for classical propagations.
Note that the implementation of \texttt{space.dof} as a \textsc{Matlab} cell vector provides some flexibility.
In multidimensional simulations, such a vector can comprise objects of different classes thus allowing the use of different DVR schemes for different spatial degrees of freedom. 
In those cases, \textsc{WavePacket} represents wavefunctions and operators using direct products of the respective one-dimensional grid representations.

For propagations using \texttt{qm\_propa}, also a temporal discretization has to be provided.
For the example considered here, this can be achieved by setting the following properties of object \texttt{time.steps}
{\small\begin{verbatim}
time.steps.m_start  = 000; 
time.steps.m_stop   = 100; 
time.steps.m_delta  = 0.025; 
time.steps.s_number = 500;
\end{verbatim}}
\noindent which specifies 100 time steps with a constant width of 0.025. 
After each of these main time steps, expectation values of relevant observables are calculated and output to the \textsc{Matlab} console and the logfile.
Internally, the main time steps are divided into shorter ‘‘substeps’’, here 500 each, which are  actually used as propagation steps for the short time propagators to be introduced in Secs.~\ref{sec:q_m} and \ref{sec:q_c}.

The variable \texttt{time} also contains information about the initial state. 
In the current example, the initial wave function is chosen to be a direct (outer) product of two Gaussian bell function which is realized by an object of class \texttt{gauss} (in the folder \texttt{+init}).
The parameters for the Gaussian along $R_1$ are specified by
{\small\begin{verbatim}
time.dof{1}       = init.gauss;
time.dof{1}.width = sqrt(0.005); 
time.dof{1}.pos_0 = -1/2;
time.dof{1}.mom_0 = 0;
\end{verbatim}}
\noindent where the three properties serve to specify the width parameter as well as the center of the Gaussian in position and momentum representation.
The parameters for the Gaussian along $R_2$ specified in \texttt{time.dof\{2\}} are the same as for $R_1$, except for the position which is $-1/20$.
Note that in fully classical or quantum-classical propagations, initial values for the positions and momenta are obtained by drawing normally distributed random numbers from the corresponding Wigner transform of the initial wavefunction.
  
Next, the diabatic representation of the potential energy of Eq.~(\ref{eq:JahnTeller}) is defined by the following code lines
{\small\begin{verbatim}
hamilt.coupling.n_eqs      = 3;       
for m = 1:hamilt.coupling.n_eqs  
    hamilt.pot{m,m}       = pot.taylor; 
    hamilt.pot{m,m}.hshift = [(2*m-3)/6 0];
    hamilt.pot{m,m}.coeffs = [0 0; 600 600];
    for n = m+1:3
        if n==m+1
            hamilt.pot{m,n}        = pot.taylor; 
            hamilt.pot{m,n}.coeffs = [0 100];
        else
            hamilt.pot{m,n}        = pot.taylor; 
            hamilt.pot{m,n}.coeffs = [0 0];
        end
    end
end
\end{verbatim}}
\noindent where the first line specifies the number of coupled Schr\"{o}dinger equations. 
The class \texttt{taylor} (stored in folder \texttt{+pot}) stands for a representation of the potential energy as a Taylor expansion the coefficients of which are given by class property \texttt{coeffs}. 
If required, the  point of reference can be shifted horizontally or vertically, as specified by  properties \texttt{hshift} and \texttt{vshift}, respectively.

It is emphasized that \texttt{hamilt.pot} is a \textsc{Matlab} cell matrix, thus permitting to define objects of different classes for the matrix entries which
allows for high flexibility and easy customization. 
This includes the possibility of leaving certain matrix entries empty, e.~g., when certain couplings are symmetry forbidden.
In addition to the Taylor series representation, \texttt{WavePacket} comes with a rather large choice of class definitions for frequently used model potential functions, including spline interpolation to tabulated data.
Of course, also user-supplied classes can be employed.

One of the hallmarks of the \textsc{WavePacket} software package is its ability to create graphical output ‘on the fly’, i.e., one movie frame is created for each main time step during a propagation using \texttt{qm\_propa}. Thus, errors may be discovered  already while the simulation is still running.
To create visualizations of (classical or quantum densities) by, e.~g. contour plots, an object of class \texttt{contour} (in the folder \texttt{+vis}) is created as follows
{\small\begin{verbatim}
plots.density           = vis.contour;
plots.density.represent = 'dvr';
\end{verbatim}}
\noindent where the second command is used to specify a representation in DVR (position space), as opposed to FBR (momentum space).
Many other properties of the contour plots can also be specified, see the Wiki documentation at SF.net.
Also note that the animation is saved as an MP4 file by default.

Besides contour plots, there are several other options to visualize densities in \textsc{WavePacket} such as curve plots, surface plots, flux plots, etc.
For densities in higher dimensions, mainly from fully classical or hybrid quantum-classical simulations, there is also the possibility to calculate and display reduced densities in each of the dimensions or in pairs thereof. 
Additionally, curve plots of all relevant expectation values versus time are created by this command
{\small\begin{verbatim}
plots.expect = vis.expect;
\end{verbatim}}
\noindent It is also possible to suppress graphical output by not creating objects named \texttt{plots.density} or \texttt{plots.expect} at all which may speed up WavePacket propagations considerably.
In that context, it is noted that graphical output can be also created in retrospect.
The WavePacket function \texttt{qm\_movie} can be used to visualize (wavefunction or trajectory) data from previous runs of \texttt{qm\_propa} provided that the following setting had been made
{\small\begin{verbatim}
state.sav_export = true;
\end{verbatim}}
\noindent Further properties \texttt{sav\_dir} and \texttt{sav\_file} serve to specify directory and file name template, respectively, for saving the data.

\section{Quantum-mechanical simulations}
\label{sec:q_m}

When providing an object of class \texttt{wave} as input argument, the \textsc{WavePacket} function \texttt{qm\_propa} numerically solves the time-dependent Schr\"{o}dinger equation (TDSE).
Internally, this is always done using a diabatic representation $\mathbf{V}$ of the light particle (typically electrons) energies
\begin{equation}
\mathbf{H}^\mathrm{dia}=\mathbf{V}(R) - \frac{1}{2M}\Delta_R \mathbf{1}
	\label{eq:ham_dia}
\end{equation}
For reasons of simplicity we have assumed only a single type of heavy particles (typically nuclei) of mass $M$; generalization to several particles with individual masses is straight-forward.
A typical example for such a diabatic Hamiltonian can be found in Eq.~(\ref{eq:JahnTeller}), see also our remarks on how to set up the (real symmetric) potential energy matrix $\mathbf V(R)$ in the previous Sec.~\ref{sec:init}.
Normally, such a diabatic matrix can be set up directly using physical/chemical model Hamiltonians for e.~g. electron-phonon coupling  \cite{Giustino2017} or molecular vibronic coupling where such an approach is sometimes referred to as ``diabatization by ansatz'' \cite{Viel2004,Eisfeld2005}.

For applications in molecular sciences, however, often an adiabatic representation is used
\begin{equation}
	\mathbf{H}^\mathrm{adi}=\mathbf{E}(R) - \frac{1}{2M}\left(\Delta_R \mathbf{1} + 2 \nabla_R \cdot \mathbf{F}(R) + \mathbf{G}(R) \right)
	\label{eq:ham_adi}
\end{equation}
where the adiabatic potential energy surfaces $E(R)$ are obtained as eigenvalues of the diabatic potential matrix V(R).
The corresponding non-adiabatic coupling (NAC) tensor elements are obtained from the adiabatic light particle (electronic) wavefunctions $\phi_i^\mathrm{adi}$ via
\begin{eqnarray}
	F^k_{ij}(R) &=& \langle \phi_i^\mathrm{adi} | \nabla_{R_k}| \phi_j^\mathrm{adi} \rangle \\
	G_{ij}(R) &=& \langle \phi_i^\mathrm{adi} | \Delta_{R}  | \phi_j^\mathrm{adi} \rangle
	\label{eq:non_adi}
\end{eqnarray}
It is well known that these quantities become very large or even diverge at (avoided or genuine) conical intersections or seams of the adiabatic potential energy surfaces, thus rendering them the main sources of non-adiabatic transitions \cite{Domcke2004,Baer2006}.

In molecular sciences, the adiabatic representation is preferred, mainly for two reasons.
First, quantum-chemical electronic structure calculations typically yield adiabatic potential energy surfaces $E(R)$, optionally also the NACs.
Second, the adiabatic representation directly allows to derive an adiabatic limit of uncoupled dynamics on each of the surfaces \cite{Panati2007}.
Nevertheless, in \textsc{WavePacket} all of the quantum--mechanical calculations are carried out within the diabatic representation in order to avoid numerical difficulties with the (near) singularities of the NACs $\mathbf{F}$ and/or $\mathbf{G}$~\footnote{Under certain circumstances a diabatization of  adiabatic data is possible \cite{Domcke2004,Baer2006}.
However, such a procedure is not trivial and outside the scope of the \textsc{WavePacket} software.}.
However, \textsc{WavePacket} offers the possibility to transform quantum-dynamical simulations in retrospect from a diabatic to the adiabatic representation by issuing the following \textsc{Matlab} command lines  
{\small\begin{verbatim}
hamilt.coupling.represent  = 'adi';
hamilt.coupling.ini_rep    = 'adi';
hamilt.coupling.ini_coeffs = [0 1 0];
\end{verbatim}}
\noindent The first line specifies an adiabatic representation to be used; the second line indicates that also the initial data refers to the adiabatic picture.
The initial data itself is given in the third line. Here, all of the density is initially set to be in the second (i.~e. first excited) adiabatic state.
As a result of these settings, all of the \textsc{WavePacket} ouput, i.~e. both the expectation values and (animated) densities is transformed to adiabatic representation.
In the absence of the above settings in the \texttt{qm\_init} file, the default is to skip these transformations and to give all output in diabatic representation.

In its numerical approach to quantum dynamics, \textsc{WavePacket} expands all wave functions and relevant operators in finite basis representations (FBRs) and/or associated discrete variable representations (DVRs), see Ref.~\cite{Light2000} as well as Sec.~III C of Part~I. 
The specification of the DVR schemes with their parameters are contained in cell vector \texttt{space.dof} as explained in Sec.~\ref{sec:init}. 
Note that they include the masses (and potentially other parameters of the associated kinetic operator), see Sec.~\ref{sec:init}.
The temporal discretization defined in object \texttt{time.steps} has to be complemented by the choice of a suitable numerical propagation scheme~\cite{Leforestier1991}.
For example, a second order differencing scheme~\cite{Askar1978} can be utilized by creating an object of class \texttt{differencing} (from folder \texttt{+tmp})
{\small\begin{verbatim}
time.propa = tmp.differencing;
time.propa.order = 2;
\end{verbatim}}
\noindent Alternatively, class \texttt{splitting} implements split operator schemes where setting the error order to 1 or 2 invokes Lie-Trotter or Strang-Marchuk splitting methods, respectively~\cite{Fleck1976,Feit1982}.
While both differencing and splitting propagators require rather short time steps, WavePacket also offers a polynomial propagator where the time
evolution operator is expanded in a truncated series of Chebychev polynomials~\cite{Tal-Ezer1984}.
Allowing for a much longer time step, this propagator is known be fast and highly accurate at the same time.
Because the efficiency of the Chebychev scheme depends on the spectral range of the Hamiltonian being not too large, the following (optional) settings may be advisable 
{\small\begin{verbatim}
hamilt.truncate.e_min = -100;
hamilt.truncate.e_max = +500;
\end{verbatim}}
\noindent This serves to truncate the grid representations of kinetic and potential energies at the given values.

\section{Quantum-classical simulations}
\label{sec:q_c}

When providing an object of class \texttt{traj} as input argument, the \textsc{WavePacket} function \texttt{qm\_propa} numerically solves the mixed (or hybrid) quantum-classical Liouville equation (QCLE) which can be derived from fully quantum-mechanical dynamics in the following way.
First, a quantum Liouville--von Neumann equation is set up for the matrix-valued Hamiltonians of Eqs.~(\ref{eq:ham_dia}) or (\ref{eq:ham_adi}). 
Then a (partial) Wigner transform is carried out with respect to the heavy particle positions $R$ and momenta $P$ only~\cite{Kapral1999,Horenko2002b}.
Finally, quantum-classical dynamics is obtained as a first order approximation in the smallness parameter $\epsilon\equiv\sqrt{m/M}$ derived from the mass ratio of light ($m$) and heavy ($M$) particles, typically electrons and nuclei~\cite{Horenko2002b}.
The resulting QCLE governing the evolution of matrix-valued phase space densities $\mathbf{X}(R,P,t)$ in the diabatic representation is given by
\begin{eqnarray}
\partial_t \mathbf{X}_W^{\rm dia}(R,P,t) &=& 
-i\left[
\mathbf{V}(R), \mathbf{X}_W^{\rm dia}(R,P,t)
\right]_-
\nonumber \\ 
&&- \frac{P}{M} \cdot \nabla_R \mathbf{X}_W^{\rm dia}(R,P,t) \nonumber \\
&&+\frac{1}{2}
\left[
\nabla_R \mathbf{V}(R), \nabla_P \mathbf{X}_W^{\rm dia}(R,P,t)
\right]_+
\label{eq:qcle_dia}
\end{eqnarray}
where $[\cdot,\cdot]_-$ and $[\cdot,\cdot]_+$ stand for commutators and  anticommutators, respectively, and $\mathbf{V}(R)$ is the diabatic potential energy matrix.
Note that this equation exactly reproduces full quantum dynamics for the special case of the potential and kinetic operators being second order polynomials such as in Eq.~(\ref{eq:JahnTeller}). 
Alternatively, an adiabatic formulation of the QCLE can be derived from Eq.~(\ref{eq:ham_adi})
\begin{eqnarray}
\partial_t \mathbf{X}_W^{\rm adi}(R,P,t) &=& 
-i \left[
\mathbf{E}(R) - i \frac{P}{M} \cdot \mathbf{F}(R), \mathbf{X}_W^{\rm adi}(R,P,t)
\right]_-\nonumber \\
&&+\frac{1}{2}\left[
\mathbf{E}(R),
\left[
\mathbf{F}(R),\nabla_P \mathbf{X}_W^{\rm adi}(R,P,t)
\right]_+
\right]_-
\nonumber \\ 
&&- \frac{P}{M} \cdot \nabla_R \mathbf{X}_W^{\rm adi}(R,P,t)\nonumber \\
&&+\frac{1}{2}
\left[
\nabla_R \mathbf{E}(R), \nabla_P \mathbf{X}_W^{\rm adi}(R,P,t)
\right]_+
\label{eq:qcle_adi}
\end{eqnarray}
where $\mathbf{E}(R)$ and $\mathbf{F}(R)$ stand for the (diagonal) adiabatic potential energy matrix and the (off-diagonal) first order NAC vectors, see Sec.~\ref{sec:q_m}.

The well-known surface hopping trajectory (SHT) schemes which were originally derived empirically~\cite{Tully1971,Tully1990} can be viewed as the simplest approaches to a numerical solution of the (diabatic or adiabatic) QCLE.
They are based on swarms of point particles representing the phase-space densities (diagonal entries of $\mathbf{X}$).
While evolving classically along the (diabatic, $\mathbf{V}(R)$, or adiabatic, $\mathbf{E}(R)$) potential energy surfaces, these trajectories may stochastically hop between the surfaces according to probabilities derived from the quantum nature of the system.
In the present version of the \textsc{WavePacket} software package, SHT schemes can be invoked as follows 
{\small\begin{verbatim}
time.hop = hop.fssh;       
\end{verbatim}}
\noindent In this example, the field \texttt{time.hop} becomes an object of the class \texttt{fssh} (inside package folder \texttt{+hop}) which represents an implementation of the fewest switches surface hopping (FSSH) technique. 
For the details of this approach, as well as the other three variants currently implemented (\texttt{mssh, lz\_1, lz\_2}), see Secs.~\ref{sec:q_c_fssh} and \ref{sec:q_c_sssh} below.
When the field \texttt{time.hop} is not initialized, surface hopping is disabled and the dynamics is purely classical, see the following Sec.~\ref{sec:q_c_class}.

Already in the early works on SHT techniques there was the idea of conserving the energy when a trajectory is hopping between two states.
This can be achieved by scaling up the momenta upon a transition from a higher to a lower potential energy surface. 
Vice versa, scaling down the momenta  when jumping to a higher potential energy surface is not always possible without violating energy conservation which leads to ``frustrated hops''.
In \textsc{WavePacket} this rescaling of the momenta is activated by the following setting
{\small\begin{verbatim}
time.hop.rescale = true;                    
\end{verbatim}}
\noindent While originally introduced empirically, at least for the adiabatic formulation this rescaling can be justified theoretically from the QCLE.
In Refs.~\cite{Kapral1999,Horenko2002b,Subotnik2013,Subotnik2016} it has been shown that it can be derived as an approximation to the non-local second term on the right-hand-side of Eq.~(\ref{eq:qcle_adi}).

Finally, in the literature there are different suggestions with respect to the direction of the momentum adjustment.
In \textsc{WavePacket}, the default is to rescale along the direction  of the momenta prior to the transition.
As an alternative, the following setting
{\small\begin{verbatim}
time.hop.sca_nac = true;                    
\end{verbatim}}
\noindent can be used to activate rescaling along the first order NAC coupling vectors $\mathbf{F}$.

\subsection{Purely classical dynamics}
\label{sec:q_c_class}
When in trajectory simulations using \textsc{WavePacket} software the field \texttt{time.hop} is not initialized, surface hopping is disabled and the dynamics is purely classical, i.~e. trajectories are propagated without undergoing any transitions. 
In that case, the trajectories are following diabatic or adiabatic potential energy surfaces, depending on the setting 'dia' or 'adi' in field \texttt{hamilt.coupling.represent} introduced in Sec.~\ref{sec:q_m}. 
In the former case, calculation of the underlying forces as the negative gradients of the diagonal element of the diabatic potential matrix $\mathbf{V}(R)$ is straight-forward.
In the latter case, however, calculating the adiabatic forces is more demanding, see App.~\ref{sec:appA}.

The choice of the initial population of (diabatic or adiabatic) states is governed by the setting of \texttt{hamilt.coupling.ini\_rep}.
If properties \texttt{represent} and \texttt{ini\_rep} of object \texttt{hamilt.coupling} are chosen equal, distributing the trajectories among the states is straightforward, using probabilities obtained from the squares of the entries of \texttt{ini\_coeffs}, see also Sec.~\ref{sec:q_m}.
Else, this coefficient vector has to be transformed from diabatic to adiabatic representation or vice versa.  

For the actual propagation of the classical trajectories, \textsc{WavePacket} offers a choice of two classes, in analogy to the short time propagators used for propagations of wavefunctions, compare also Sec.~\ref{sec:q_m}: 
For example, with the following settings 
{\small\begin{verbatim}
time.propa = tmp.differencing;
time.propa.order = 3;
\end{verbatim}}
\noindent a Stoermer-Verlet integrator is invoked~\cite{Verlet1967,Frenkel2002} which is the classical equivalent to the second order differencing in quantum dynamics. 
Alternatives are integrators based on Trotter (first order) or Strang (second order) splitting approaches which are invoked by specifying \texttt{time.propa = tmp.splitting}. 
The latter one corresponds to the ``leap frog'' integrator commonly used in classical molecular dynamics~\cite{Frenkel2002}. 
Finally, also Beeman's third order algorithm ~\cite{Beeman1976} and Yoshida's fourth order algorithm ~\cite{Yoshida1990} have been implemented.
Note that these propagators are also used for all SHT simulations in between non-adiabatic transitions.

\subsection{Fewest switches surface hopping}
\label{sec:q_c_fssh}

In the first family of SHT simulation techniques presented here, a quantum state vector (or a density matrix) is followed for each of the trajectories to reflect the quantum nature of the coupled state problem.
Using the diabatic representation of Eq.~(\ref{eq:ham_dia}), the evolution of the corresponding coefficient vector $\mathbf{c}$ is governed by
\begin{equation}
	i\dot{\mathbf{c}} = \mathbf{V} \mathbf{c}
	\label{eq:tdse_dia}
\end{equation}
where the right-hand-side is time-dependent because the potential energy matrix $V$ is evaluated for the positions of the classical particles $R(t)$.
Alternatively, the adiabatic representation of Eq.~(\ref{eq:ham_adi}) can be used
\begin{equation}
	i\dot{\mathbf{c}} = \left(\mathbf{E}-i\frac{P}{M}\cdot \mathbf{F}\right) \mathbf{c}
	\label{eq:tdse_adi}
\end{equation}
where $P$ and $M$ stand for the momenta and masses of the classical particles, respectively.
For the calculation of the first-order NAC coupling vectors $\mathbf{F}$ from the diabatic potential matrix $\mathbf{V}$ in \textsc{WavePacket}, see App.~\ref{sec:appA}.

The simplest way of obtaining quantum-based probabilities for hopping from state $m$ to state $n$ will be simply to use the density in the target state itself~\cite{Tully1971}
\begin{equation}
	\gamma_{m\rightarrow n} = \rho_{nn}
	\label{eq:mssh}
\end{equation}
where we have introduced the usual density notation $\rho_{mn}=c_m^\ast c_n$.
In principle, this algorithm will produce the correct populations for a large enough ensemble of trajectories. 
In practice, however, there will be many hopping events at all times, even when the trajectories are outside the transition regions.
Hence, this algorithm will be termed ``multiple switches surface hopping'' (MSSH) in our  \textsc{WavePacket} implementation.
It is invoked by the following command 
{\small\begin{verbatim}
time.hop = hop.mssh;       
\end{verbatim}} \noindent
The rapid switching behavior renders this method inferior to any of the other SHT variants presented here; it is included here only for reasons of historical completeness.

The fewest switches surface hopping (FSSH) algorithm represents a substantial improvement over the MSSH algorithm. Because of its simplicity, it has gained enormous popularity since its first publication in 1990~\cite{Tully1990}.
In FSSH, the hopping probability from state $m$ to state $n$ is based on the rate of change $d\rho_{nn}/dt$ of the density of the target state.
In a diabatic picture this probability is given by 
\begin{equation}
	\gamma_{m\rightarrow n} = \frac{2 \Delta t}{\rho_{mm}} \Im (\rho_{nm} V_{nm})
	\label{eq:fssh_dia}
\end{equation}
where $\Delta t$ stands for the time step size.
Alternatively, in an adiabatic picture this probability amounts to
\begin{equation}
	\gamma_{m\rightarrow n} = \frac{2 \Delta t}{\rho_{mm}} \Re (\rho_{nm} \frac{P}{M}\cdot F_{nm})
	\label{eq:fssh_adi}
\end{equation}
Indeed, SHT simulations using these two formulae can be shown to minimize the number of state switches, subject to maintaining the correct statistical distribution of the populations at all times~\cite{Tully1990}.
In \textsc{WavePacket}, these FSSH algorithms are invoked by the following command
{\small\begin{verbatim}
time.hop = hop.fssh;       
\end{verbatim}} \noindent
Note that here and throughout the following, all hopping probabilities $\gamma$ are truncated such as to be bounded inside $[0,1]$.

\begin{figure}
\begin{centering}
\includegraphics[width=0.8\textwidth]{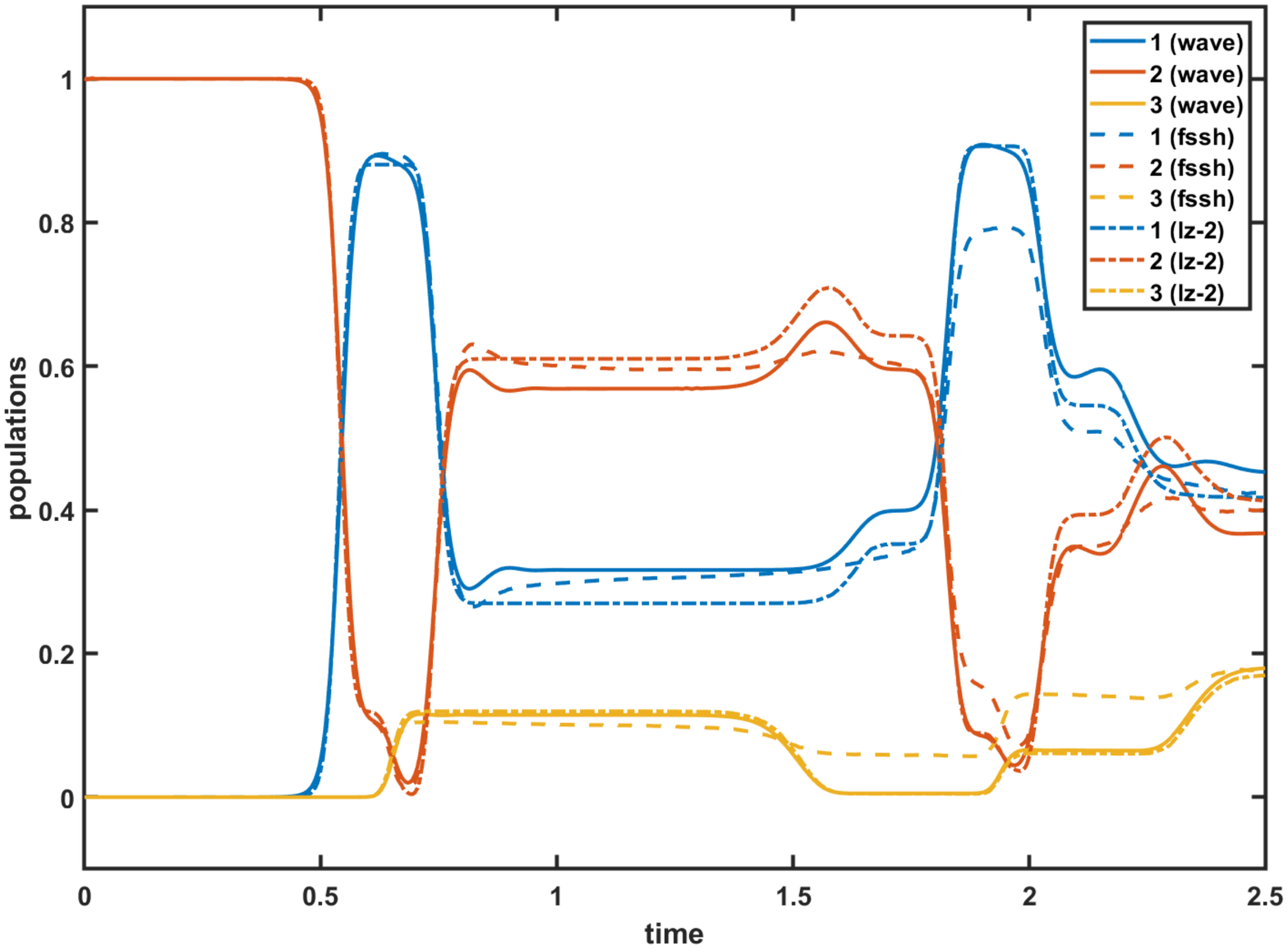}
\end{centering}
\caption{Population dynamics for the Hamiltonian of Eq. (\ref{eq:JahnTeller}): Population of the three adiabatic states (numbered according to ascending energy) for fully quantum-mechanical (full curves) versus quantum-mechanical propagations. Fewest switches surface hopping (dashed curves) and single switch surface hopping (dash-dotted curves). 
Generated with \textsc{WavePacket} option: \texttt{plots.expect = vis.expect}}
\label{fig:Populations}
\end{figure}

As an example we show in Fig.~\ref{fig:Populations} the population transfer for the system introduced in Sec.~\ref{sec:init}, comparing numerically exact quantum dynamics with SHT approximations. 
While the FSSH algorithm reproduces the first population transfer ($2\rightarrow 1$) practically exactly, there are minor deviations of the populations after the second transfer ($2\rightarrow 1$).
A closer inspection shows that these discrepancies are due to geometric phase effects.
Nonetheless, also the shape of the (here not very pronounced) Stueckelberg oscillations is at least qualitatively reproduced. 

\subsection{Single switch surface hopping}
\label{sec:q_c_sssh}

The second family of SHT approaches implemented in \textsc{WavePacket} is not requiring integration of quantum state vectors along each of trajectory.
These SHT algorithms are referred to as ``single switch surface hopping'' (SSSH) because the hopping probability is only evaluated once for every passage of transition regions, typically intersections of adiabatic potential energy surfaces.
Assuming a locally linear double cone topology of the surfaces in these regions, the transition probabilities in SSSH algorithms are based on variants of Landau-Zener (LZ) formulae~\cite {Lasser2007,Lasser2008,Fermanian2008,Belyaev2011,Belyaev2014,Belyaev2015}.

The first single switch variant implemented in \textsc{WavePacket} is accessed by
{\small\begin{verbatim}
time.hop = hop.lz_1;       
\end{verbatim}} \noindent
Essentially, it represents the conventional analytic LZ result.
In a diabatic representation, the probability for hopping from state $m$ to state $n$ is given by
\begin{equation}
	\gamma_{m\rightarrow n} = \exp \left( - 2\pi \frac{ V_{nm}^2}{\frac{d}{dt} |V_{nn}-V_{mm}|} \right)
	\label{eq:lz2_dia}
\end{equation}
where the time-derivative of the diabatic energy gap is obtained by finite differencing along the trajectories.
Note that the formula is evaluated only at the center of a nonadiabatic region, i.~e. where diabatic potentials intersect each other.
In an adiabatic formulation, the hopping probability yields \cite{Fermanian2008,Belyaev2011}
\begin{equation}
	\gamma_{m\rightarrow n} = \exp \left( - \frac{\pi}{4}\frac{Z_{nm}}{|\frac{P}{M}\cdot F_{nm}|} \right)
	\label{eq:lz1_adi}
\end{equation}
Also this formula is applied only once for each nonadiabatic region, namely whenever an eigenvalue gap $Z_{nm}\equiv|E_n-E_m|$ becomes minimal along an individual classical trajectory.

The necessity to calculate NAC vectors $\mathbf{F}$ is circumvented elegantly in the second single switch variant implemented in \textsc{WavePacket} which can be invoked by
{\small\begin{verbatim}
time.hop = hop.lz_2;       
\end{verbatim}} \noindent
This approach is available in an adiabatic picture only, and the probability for surface hopping is expressed only in terms of adiabatic energy gaps and second time derivatives thereof \cite{Belyaev2011,Belyaev2014,Belyaev2015}
\begin{equation}
	\gamma_{m\rightarrow n} = \exp \left( - \frac{\pi}{2}\sqrt{\frac{Z_{nm}^3}{\frac{d^2}{dt^2}Z_{nm}}}\right)
	\label{eq:lz2_adi}
\end{equation}
which again is evaluated only at local minima of energy gaps $Z_{nm}$.
This formulation offers the unique advantage of not requiring nonadiabatic coupling information any more, which makes this method  not only more efficient than any of the above methods but it is also in line with the application of electronic structure methods in molecular sciences where often only adiabatic energy surfaces $\mathbf{E}$ are available. 

The accuracy of the SHT algorithm based on Eq.~(\ref{eq:lz2_adi}) is also shown in Fig.~\ref{fig:Populations}.
For the test system of Sec.~\ref{sec:init}, SSSH and FSSH reproduce the population transfer from numerically exact quantum dynamics with roughly equal quality.
However, the weak Stueckelberg oscillation structure is absent in the SSSH results, due to the semi-classical nature of the underlying LZ approximation.  
\section{Summary and outlook}
\label{sec:summary}

The present version 6.0.2 of \textsc{WavePacket} represents a major step forward from previous versions 5.x of that software, with the main new feature being the addition of classical trajectories and surface hopping techniques.
To the best of our knowledge, \textsc{WavePacket} is the only software package that can be used for fully classical,  mixed quantum-classical, and fully quantum-mechanical propagations of physical/chemical systems on an equal footing, i.~e., for the same Hamiltonian, initial conditions and time stepping.
On the one hand, for low--dimensional systems this allows for a direct comparison of classical versus quantum dynamics which can be used, e.~g., to identify quantum effects and assess their importance for various simulation tasks.
On the other hand, the quantum-classical propagation techniques allow to substantially increase the number of degrees of freedom that can be treated with \textsc{WavePacket}, at least for systems with a clear separation of fast and slow coordinates where quantum effects can be restricted to the former ones.

Technically, these changes of \textsc{WavePacket} have been made possible by a major rewrite  in an object-oriented manner.
While the oldest versions of our software package were still written in a relatively traditional, completely procedural way, the introduction of generalized DVR/FBR methods in version 4.5 led us to make (limited) use of the object-oriented features offered by \textsc{Matlab}.
These approaches have now been extended to large parts of the code of version 6.0.2.
In particular, by introducing the \textsc{WavePacket} main classes \texttt{wave} and \texttt{traj}, the function \texttt{qm\_propa} has been made polymorph, i.~e., being able to propagate quantum and classical objects alike.
Along these lines, further extensions of the software package will be relatively easy to realize.
This includes the main classes \texttt{ket} and \texttt{rho} for the implementation of quantum state vectors and density matrices in eigen representation which are currently under development.
Future releases of \textsc{WavePacket} will also include class definitions implementing various semi-classical approaches based on Gaussian packets in phase space~\cite{Horenko2002a,Lasser2017,Buchholz2018}. 
Of particular interest will be nonadiabatic extensions of Gaussian propagation methods, such as the multiple spawning technique~\cite{BenNun2002}, surface hopping Gaussian propagations \cite{Horenko2002b}, or adaptive variants thereof \cite{Horenko2004}. 

In addition to the \textsc{WavePacket} main classes described above, also many other parts of our software package are now based on \textsc{Matlab} classes.
Currently there are almost 100 class definitions which are used especially where choices are  to be made for a user-defined setup of the system to be simulated.
As has been shown in Sec.~\ref{sec:init}, these options encompass DVRs/FBRs, kinetic and potential energy operators, initial states, propagators, visualization types, etc..
This also includes the choice of different types of SHT techniques which have recently been added to our software.
It is planned to add more variants here to keep up with the recent progress in surface hopping \cite{Wang2016}.
In summary, the  introduction of object-oriented techniques has been instrumental in achieving a fully modular design, thus making the codes much more flexible, in order  to cover the growing diversity of physical/chemical systems that can be simulated with \textsc{WavePacket}.

\begin{acknowledgments}
This work has been supported by the Einstein Center for Mathematics Berlin (ECMath) through project SE~20 and by the Berlin Mathematics Research Center MATH$+$ through project AA2--2.
Caroline Lasser is acknowledged for insightful discussions on surface hopping trajectories.
We are grateful to Ulf Lorenz (formerly at U Potsdam) for his valuable help with all kind of questions around the \textsc{WavePacket} software package.
\end{acknowledgments}

\begin{appendix}
\section{Computation of adiabatic forces and NAC vectors}
\label{sec:appA}

In order to perform classical or quantum-classical dynamics in the adiabatic picture, the gradients of the adiabatic potentials are necessary. 
After diagonalizing the diabatic potential matrix
\begin{equation}
\mathbf{E}(R) = \mathbf{U}(R)^T \mathbf{V}(R) \mathbf{U}(R)
\label{eq:A_diag}
\end{equation}
and storing its eigenvectors $u_i(R)$, \textsc{WavePacket} evaluates the gradients of the adiabatic potential energy surfaces by applying the Hellmann-Feynman theorem
\begin{equation}
\nabla_{R_k} E_i(R) = u_i(R)^T \left( \nabla_{R_k} \mathbf{V}(R) \right) u_i(R)
\label{eq:A_adi}
\end{equation}
where the knowledge of the gradient of the diabatic potential matrix is required.

For the adiabatic variant of the FSSH algorithm, see Eq.~(\ref{eq:fssh_adi}), the hopping probability depends on the NAC vectors. 
By using the eigenvectors of the diabatic potential matrix, these vectors can be represented by 
\begin{equation}
F^k_{ij}(R) = u_i(R)^T \nabla_{R_k} u_j(R).
\end{equation}
However, this representation depends on the gradients of the eigenvectors which, typically, are not available. 
To avoid this problem, \textsc{WavePacket} uses the following formula for the NAC vectors
\begin{equation}
F^k_{ij}(R) = \frac{ u_i(R)^T \left( \nabla_{R_k} \mathbf{V}(R) \right) u_j(R) } {E_j(R) - E_i(R)}
\label{eq:A_nac}
\end{equation}
where, again, the knowledge of the gradient of the diabatic potential matrix is required.

\begin{figure}
\begin{centering}
\includegraphics[width=0.6\textwidth]{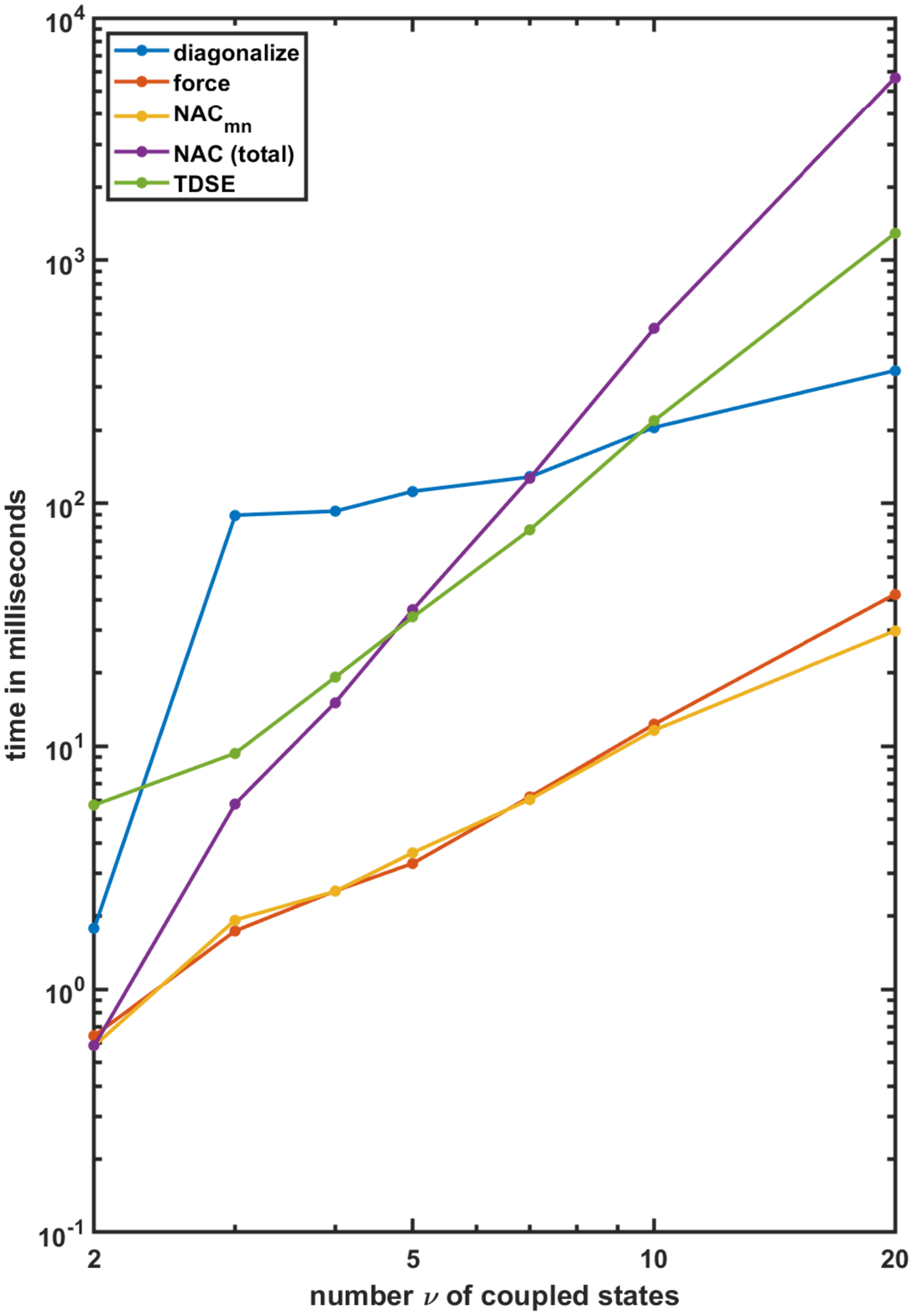}
\end{centering}
\caption{Double-logarithmic representation of \textsc{WavePacket} computational time for diagonalization of $\mathbf{V}$ and calculation of adiabatic forces as well as of NAC vectors $\mathbf{F}$ in FSSH for 2 dimensions and varying number $\nu$ of coupled channels.
For 100 time steps of an FSSH simulation with 1000 trajectories in adiabatic representation.}
\label{fig:Runtimes}
\end{figure}

Average computation times of \textsc{WavePacket} on a Macbook  computer (1.3 GHz Intel Core i5 processor; 4 GB 1600 MHz DDR3 memory) are shown in Fig.~\ref{fig:Runtimes}.
The measured times are for the case of adiabatic FSSH simulations in two dimensions, generalizing the example defined in Sec.~\ref{sec:init} to a variable number $\nu$ of coupled states.
First, we note that for $\nu=2$, the diagonalizations as well as the calculations of forces and NAC vectors are carried out analytically which is the reason why these calculations are very fast.
Hence, we will consider only the data for $\nu>2$ in the following.
There, the time for diagonalization (\ref{eq:A_diag}) rises only very slowly with increasing $\nu$ which is due to the sparsity of matrix $\mathbf{V}$ for our example.
The other curves, however, behave as expected, i.~e., the effort to calculate adiabatic forces (\ref{eq:A_adi}) as well as $\mathbf{F}$ vectors (\ref{eq:A_nac}) for a single pair of adiabatic states scales as $\mathcal{O}(\nu^2)$.
Consequently, the cost to calculate the whole matrix of first order NAC vectors scales  
as $\mathcal{O}(\nu^4)$.
Also shown is the time used for the numerical TDSE integrator (based on diagonalization of  $\mathbf{V}$) which scales as $\mathcal{O}(\nu^3)$.
Note that the latter two contributions are omitted in SSSH simulations which makes them considerably faster than FSSH for approximately $\nu>10$.

\end{appendix}

\bibliography{WavePacket3}

\clearpage

\clearpage

\clearpage

\end{document}